 \documentclass[twocolumn,prb,showpacs,amsmath,amssymb,floatfix]{revtex4}

\usepackage{graphicx}
\usepackage{dcolumn}
\usepackage{bm}
\usepackage{xspace}

\newcommand{\eg}{e.\,g.\@\xspace}   
\newcommand{\tetal}{\emph{et.\@\xspace al.\@\xspace}}

\newcommand{\tref}{Ref.\@\xspace}   
\newcommand{\trefs}{Refs.\@\xspace} 

\newcommand{\teq}{Eq.\@\xspace}     

\newcommand{\tfig}{Fig.\@\xspace}   

\newcommand{\tchap}{Sect.\xspace}      

\newcommand{\onehalf}{1/2}
\newcommand{\oh}{\onehalf}
\newcommand{\spinonehalf}{S\!=\!\oh}

\begin{document}
\setlength{\unitlength}{1mm}
\title{Thermodynamic properties of the two-dimensional $\spinonehalf$ Heisenberg antiferromagnet coupled to bond phonons}

\author{Carsten H.~Aits}%
\email{ca@thp.uni-koeln.de}%
\affiliation{Institut f\"ur Theoretische Physik, Universit\"at zu K\"oln, Z\"ulpicher Str.~77, D-50937 K\"oln, Germany.}

\author{Ute L\"ow}%
\email{ul@thp.uni-koeln.de}%
\affiliation{Institut f\"ur Theoretische Physik, Universit\"at zu K\"oln, Z\"ulpicher Str.~77, D-50937 K\"oln, Germany.}%

\begin{abstract}
By applying a quantum Monte Carlo procedure based on the loop algorithm we investigate thermodynamic properties of the two-dimensional antiferromagnetic $\spinonehalf$ Heisenberg model coupled to Einstein phonons on the bonds. The temperature dependence of the magnetic susceptibility, mean phonon occupation numbers and the specific heat are discussed in detail. We study the spin correlation function both in the regime of weak and strong spin phonon coupling (coupling constants $g=0.1$, $\omega=8J$ and $g=2$, $\omega=2J$, respectively). A finite size scaling analysis of the correlation length indicates that in both cases long range N\'eel order is established in the ground state.
\end{abstract}



\maketitle

\section{Introduction}\label{intro}

During the last years, there has been considerable interest in low-dimensional spin systems with spin phonon coupling. While one-dimensional (1D) models including this mechanism have been studied extensively, little is known about two-dimensional (2D) systems with spin phonon coupling. 

In 1D systems, the mechanism of spin phonon coupling is closely connected with the phenomenon of the spin Peierls transition. Theoretical understanding of this phase transition goes back to the work of Pytte,\cite{pytte74b} who showed that a three-dimensional system consisting of uniform antiferromagnetic $\spinonehalf$ Heisenberg chains is unstable towards dimerization of the chains if coupled to three-dimensional lattice vibrations. Additionally, he showed that in the adiabatic limit of small phonon frequencies the dimerized phase can be described by a statically dimerized spin model with temperature-dependent dimerization. Cross and Fisher\cite{cross79} improved on the calculation of \tref\onlinecite{pytte74b} by treating the spin part of the Hamiltonian in continuum field theory.\cite{luther75} Their calculation yielded a convincing description of the phonon softening in the limit of small spin phonon couplings. In the case of CuGeO$_3$, however, which was the first inorganic spin Peierls compound discovered,\cite{hase93a} this treatment is not sufficient. For this reason various 1D dynamical models have been investigated, taking into account the coupling of magnetic and phononic degrees of freedom explicitly. Most publications on this issue refer to two different models, both showing a quantum phase transition between a dimerized and a N\'eel ordered phase. While in the so-called difference coupling model\cite{uhrig98b,bursi99,weisse99b,sun00,raas01a} the magnetic interaction between nearest neighbours depends on the distance between neighboured sites, the bond coupling model\cite{sandv97,weisse99b,kuehne99,sandv99,trebst01,raas02} is considered to be more realistic for describing the spin phonon coupling mechanism in CuGeO$_3$.\cite{werne99,feld00}  

In the case of the 2D $\spinonehalf$ Heisenberg antiferromagnet with spin phonon coupling discussions of statically dimerized models are found in the literature.\cite{tang88,koga98,koga99,omari00,sirker02} As in one dimension, these models are thought to describe the dimerized phase of dynamical models in the adiabatic limit. In contrast to the 1D case though it is not clear how to place alternating magnetic couplings in both spatial directions on a square lattice. By comparing ground state energies of three statically dimerized models Sirker \tetal\cite{sirker02} conclude that a stair-like distortion of the lattice is the energetically favoured dimerization pattern, contradicting an older result by Tang and Hirsch\cite{tang88} who find a plaquette-like distortion. 

The aim of this paper is to investigate thermodynamic properties of the 2D Heisenberg model coupled to bond phonons. Such a model takes into account the elastic energy due to lattice distortions which is not included in statically dimerized models. Furthermore, the whole range of phonon frequencies is accessible in our treatment.   

The paper is organized as follows. In \tchap\ref{mod} we introduce the model Hamiltonian and describe our quantum Monte Carlo algorithm. In \tchap\ref{2dtd} we discuss the temperature dependence of the magnetic susceptibility, mean phonon occupation numbers and the specific heat. An analysis of the spin correlation function is found in \tchap\ref{2dgs}. Section \ref{sum} concludes with a summary.  

\section{Model Hamiltonian and Quantum Monte Carlo method}\label{mod}

The model we consider is a generalization of the 1D bond coupling model from \trefs\onlinecite{kuehne99,raas02}. The Hamiltonian reads\cite{aits02}
\begin{align}\label{ham}
H&=\frac{J}{2}\sum_{i,j=1}^{N}(\vec{\sigma}_{ij}\vec{\sigma}_{i+1,j}-1)(1+g[a_{ij}+a_{ij}^\dagger])\notag\\
&+\frac{J}{2}\sum_{i,j=1}^{N}(\vec{\sigma}_{ij}\vec{\sigma}_{i,j+1}-1)(1+g[b_{ij}+b_{ij}^\dagger])\\
&+\omega\sum_{i,j=1}^{N}(a_{ij}^\dagger a_{ij}+b_{ij}^\dagger b_{ij})\notag.
\end{align}
Here $\vec{\sigma}_{ij}$ denote Pauli spin operators at lattice site $(i,j)$ on a square lattice, while $a_{ij}$ and $a_{ij}^\dagger$ ($b_{ij}$ and $b_{ij}^\dagger$) are phonon annihilation and creation operators on the bond between site $(i,j)$ and site $(i+1,j)$ (between sites $(i,j)$ and $(i,j+1)$). Note that we assume periodic boundary conditions.   

By shifting the phonon operators according to 
\begin{equation}\label{shift}
a_{ij}\rightarrow a_{ij}+\frac{gJ}{2\omega},\qquad b_{ij}\rightarrow b_{ij}+\frac{gJ}{2\omega}
\end{equation}
and neglecting a constant energy contribution model \eqref{ham} can be mapped onto the phenomenologically more realistic Hamiltonian
\begin{align}\label{hameff}
\tilde{H}&=\frac{1}{2}\sum_{ij}(J'+g'[a_{ij}+a_{ij}^\dagger])\vec{\sigma}_{ij}\vec{\sigma}_{i+1,j}\notag\\
&+\frac{1}{2}\sum_{ij}(J'+g'[b_{ij}+b_{ij}^\dagger])\vec{\sigma}_{ij}\vec{\sigma}_{i,j+1}\\
&+\omega\sum_{ij}(a_{ij}^\dagger a_{ij}+b_{ij}^\dagger b_{ij})\notag,
\end{align}
with rescaled coupling constants $J'\equiv J(1+g^2J/\omega)$ and $g'\equiv gJ$. Note that \teq\eqref{hameff} differs from \eqref{ham} by the absence of the (unphysical) static terms \mbox{$-Jg/2(a_{ij}^\dagger + a_{ij})$},\mbox{$-Jg/2(b_{ij}^\dagger + b_{ij})$}. Model \eqref{hameff}, however, cannot be analyzed directly in a quantum Monte Carlo study because a sign problem occurs. For this reason we choose Hamiltonian \eqref{ham} as a starting point for our analysis and shift the numerical results if needed. 

To study the properties of model \eqref{ham}, we developed a quantum Monte Carlo\cite{suzuki77} algorithm similar to the algorithm described in \tref\onlinecite{kuehne99}. First, the partition function of the 2D quantum system \eqref{ham} is mapped onto a three-dimensional classical system. Technically, this is done by means of a Trotter Suzuki decomposition.\cite{kuehne99} We then apply an update procedure which treats spin and phononic degrees of freedom separately. For the spin updates, we make use of a modified loop algorithm\cite{evertz93,evertz97} for quantum spin systems. The main advantage of the loop algorithm is that it allows global spin updates, substantially reducing autocorrelation times. Furthermore, so-called improved estimators can be used in the evaluation of the magnetic susceptibility and spin correlations. To modify the phonon occupation numbers we apply local heat bath updates. By building clusters of phonons in imaginary time direction we extended the algorithm from \tref\onlinecite{kuehne99}, diminishing autocorrelation effects even more. Obviously the detailed balance condition is fulfilled for both steps separately and thus for the whole procedure.

For the phonon updates we had to introduce a cutoff, allowing occupation numbers up to 40 phonons per bond. The effect of such a truncation of the Hilbert space is negligible if the measured mean phonon occupation numbers are more than an order of magnitude smaller than the cutoff. In \tchap\ref{nmed} we show explicitly that this condition is fulfilled. In order to take into account the high dimension of the phonon subspace of the Hilbert space, we employed the importance sampling technique and made 30 phonon updates per spin update, using only the last configuration for the evaluation of expectation values. Additionally, for each temperature the first 25\% of the sweeps were skipped for thermalization.

For Monte Carlo simulations based on a Trotter Suzuki decomposition the estimates of thermodynamic quantities depend on the inverse Trotter number squared.\cite{suzuki85} In the following sections we give the explicit value for the Trotter number $M$. With the values for $M$ chosen, we find the statistical fluctuations of our results larger than the effect of the finite Trotter number.  

Before discussing the results in detail we add one further remark concerning statistical errors. In our calculations we neglected autocorrelation effects and -- as a rough estimate -- calculated root-mean-squared errors only. If no error bars in the plots of this paper are shown, the errors are smaller than the symbol size used.

\section{Thermodynamic properties}\label{2dtd}

In this section we discuss the finite temperature properties of model \eqref{ham}. We expect that the knowledge of how a non-vanishing spin phonon coupling influences the thermodynamic properties will be of importance for the interpretation of experiments. This might be of particular interest for substances which display \eg acoustic anomalies or for which it is known that the exchange integral depends sensitively on the positions of the ions. 

Here, we confine ourselves to temperatures \mbox{$0.5J\leq T\leq 4J$}. In this temperature range, we find that the dependence of measured quantities on the system size is negligible if we consider linear system sizes $N\geq 12$. All results presented in this section were calculated on a lattice with 12$\times$12 sites, providing statements about system properties in the thermodynamic limit. If not stated differently, at each temperature 10$^5$ spin updates were executed. For the Trotter number a value of $M=80$ was chosen.

\subsection{Magnetic susceptibility}\label{sus}

\begin{figure}[t!]
\begin{center}
  \includegraphics[angle=270,width=\linewidth]{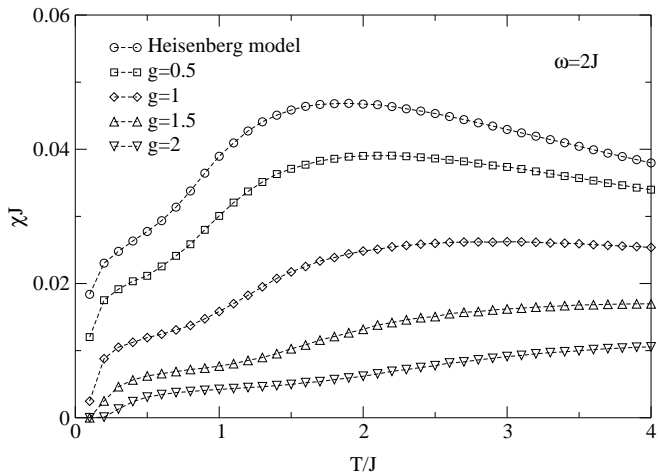}
\end{center}
\caption{Magnetic susceptibility vs.~temperature for fixed \mbox{$\omega=2J$} and spin phonon coupling $g$ between 0.5 and 2.0. For comparison Monte Carlo results for the Heisenberg model are plotted.\label{susg}}
\end{figure}
\begin{figure}[t!]
\begin{center}
  \includegraphics[angle=270,width=\linewidth]{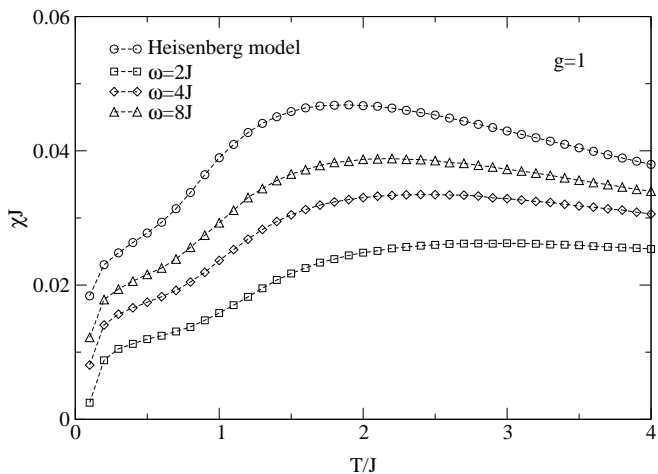}
\end{center}
\caption{Magnetic susceptibility vs.~temperature for fixed \mbox{$g=1$} and phonon frequencies $\omega$ between $2J$ and $8J$. Again for comparison Monte Carlo results for the Heisenberg model are shown.\label{susw}}
\end{figure}
We start with a discussion of the magnetic susceptibility per site $\chi$ for vanishing magnetic fields. Figure \ref{susg} shows the dependence of the susceptibility on the spin phonon coupling $g$ for fixed phonon frequency $\omega$, \tfig\ref{susw} the dependence on $\omega$ for fixed $g$. In both figures Monte Carlo results for the susceptibility of the 2D Heisenberg model are included. 

The results can be summarized as follows. For fixed $\omega$, the overall height of the susceptibility is diminished with increasing spin phonon coupling. As in the 1D case,\cite{kuehne99} a large spin phonon coupling tends to reduce the magnetic response of the system. On the other hand, for fixed $g$ the susceptibility is growing with increasing phonon frequency. In the antiadiabatic limit phononic degrees of freedom are suppressed, yielding the results of the Heisenberg model. The curves show a broad maximum which is typical for antiferromagnetic spin models. This maximum is shifted to higher temperatures with increasing $gJ/\omega$. 

\begin{figure}[t!]
\begin{center}
  \includegraphics[width=\linewidth]{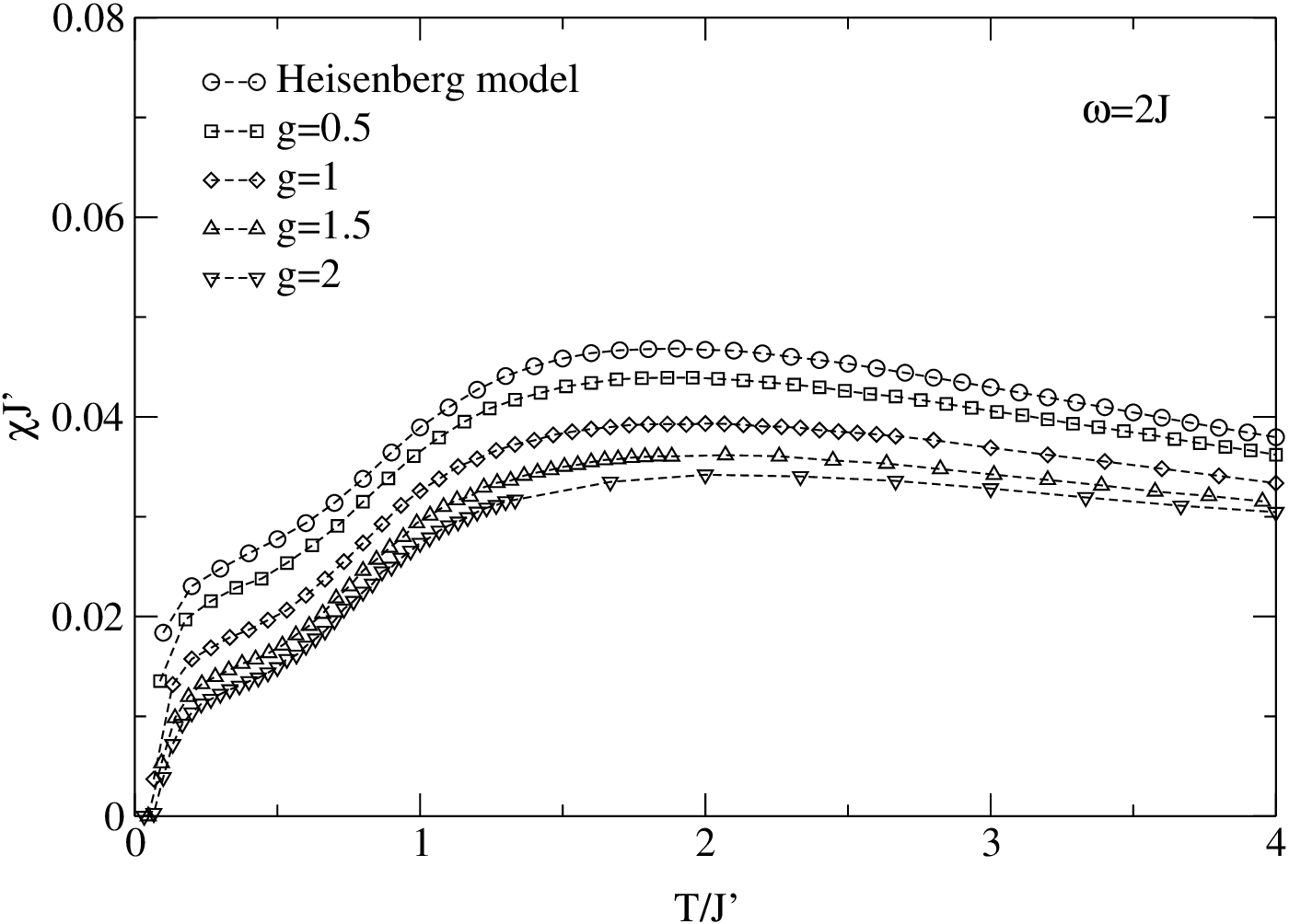}
\end{center}
\caption{The data from \tfig\ref{susg} in terms of the rescaled magnetic coupling $J'$.\label{susgres}}
\end{figure}
We find that both the shift of the position of the maximum and the reduction of magnetic response with increasing $\frac{gJ}{\omega}$ is mainly due to the static terms in \eqref{ham}. In units of the rescaled magnetic coupling $J'$ from the transformed Hamiltonian \eqref{hameff} the shift of the maximum is reduced significantly (see \tfig\ref{susgres}), and the reduction of the magnetic response due to the spin phonon coupling is not very strong. The same behaviour has already been reported for the 1D bond coupling model in \tref\onlinecite{raas02}.
\begin{figure}[t!]
\begin{center}
  \includegraphics[width=\linewidth]{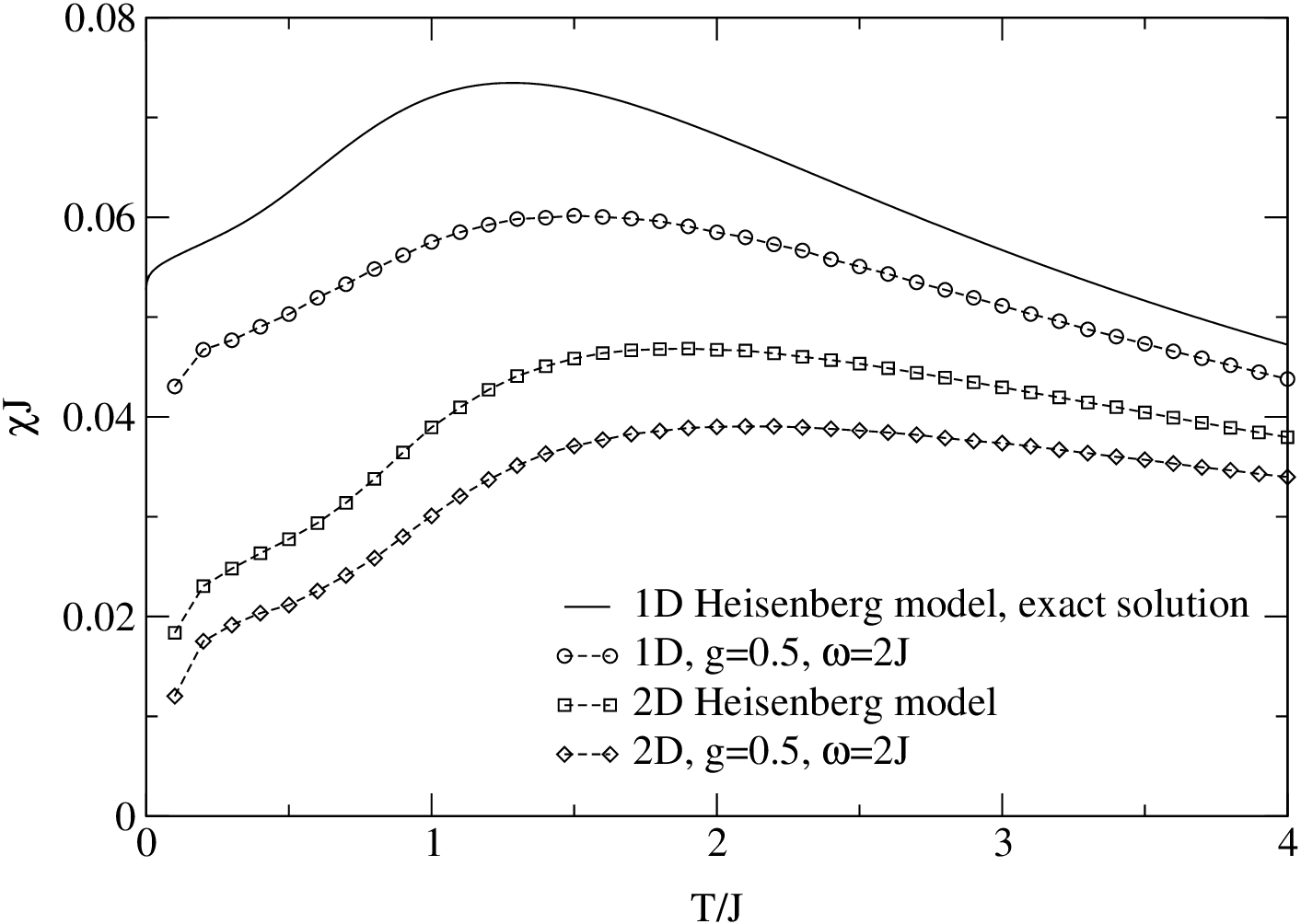}
\end{center}
\caption{Temperature dependence of the magnetic susceptibility in $d=1$ and \mbox{$d=2$} for $g=0.5$ and $\omega=2J$. For comparison in $d=1$ the exact result from \trefs\onlinecite{egger94,klump98a} and in $d=2$ Monte Carlo data for the Heisenberg model are shown.\label{sus12}}
\end{figure}

For comparison between $d=1$ and $d=2$ we return to model \eqref{ham} without phonon shift. Figure \ref{sus12} shows the susceptibilities for $g=0.5$, $\omega=2J$. Compared to the 1D case, the overall height of the susceptibility of the 2D model is diminished. This effect is explained by the larger coordination number in the 2D case, reducing the response of the system to an external magnetic field. 

Qualitatively the influence of the spin phonon coupling is similar in $d=1$ and $d=2$. In \tfig\ref{sus12} the exact result from \trefs\onlinecite{egger94,klump98a} for the 1D and the Monte Carlo results for the 2D Heisenberg model are shown. Both in 1D and 2D we find a significant shift of the maximum and a strong reduction of the maximum height as compared to the Heisenberg model.

\subsection{Mean phonon occupation numbers}\label{nmed}

\begin{figure}[b!]
\begin{center}
  \includegraphics[width=\linewidth]{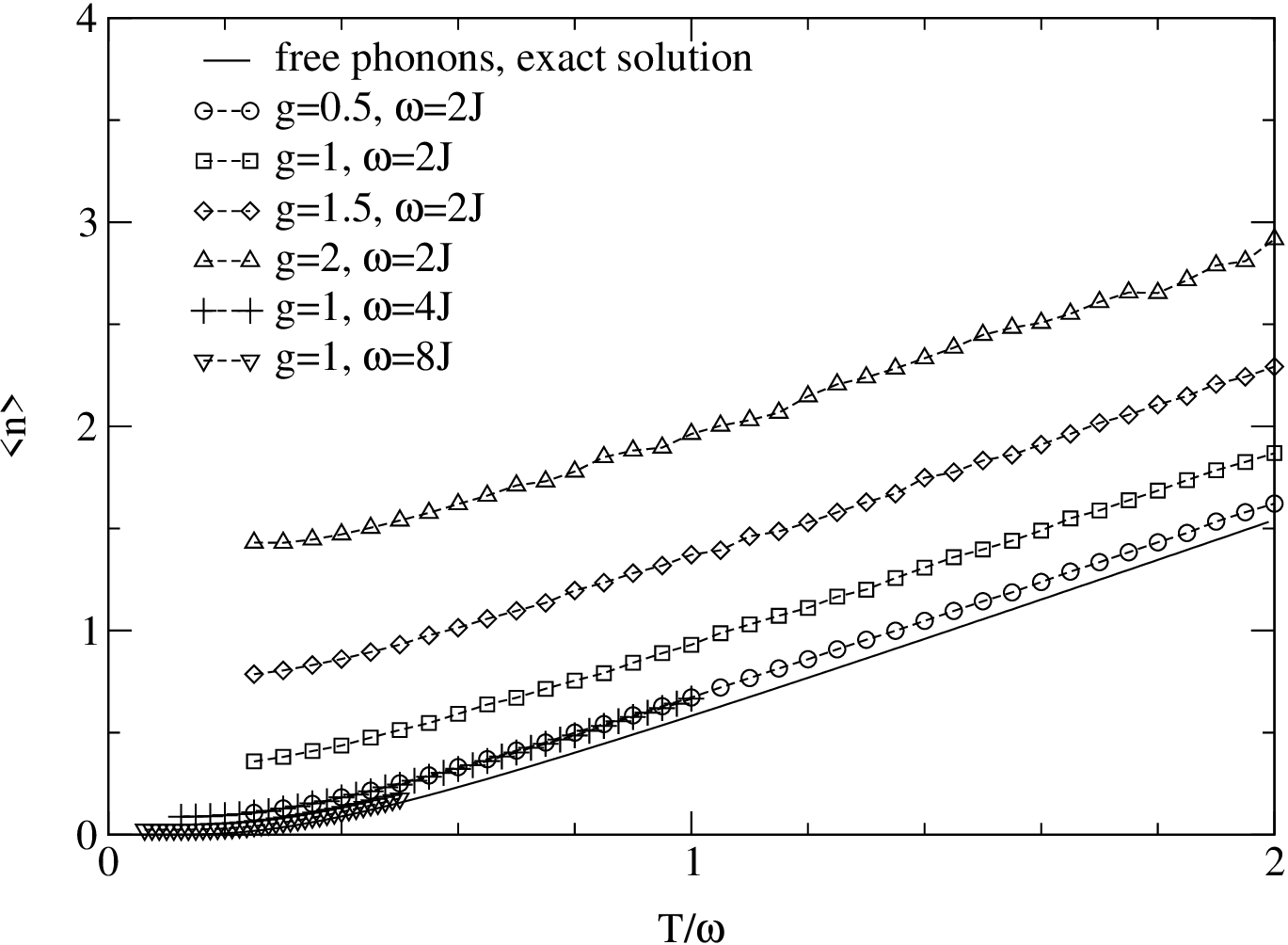}
\end{center}
\caption{Mean phonon occupation numbers $\langle n \rangle$ as a function of $T/\omega$. The solid line shows the Bose distribution $n_\text{free}$ for free Einstein phonons.\label{nga}}
\end{figure}

Further insight into the model can be gained by studying the influence of the spin phonon coupling on the mean phonon occupation numbers
\begin{align}
\langle n \rangle &=\frac{1}{N^2}\sum_{ij}\langle a_{ij}^\dagger a_{ij}\rangle\\
\langle m \rangle &=\frac{1}{N^2}\sum_{ij}\langle b_{ij}^\dagger b_{ij}\rangle.
\end{align}
as compared to the free phonon case. These numbers can be viewed as a measure of the strength of lattice vibrations and therefore allow to analyze how the lattice is influenced by the spin degrees of freedom. 

\begin{figure}[t!]
\begin{center}
  \includegraphics[width=\linewidth]{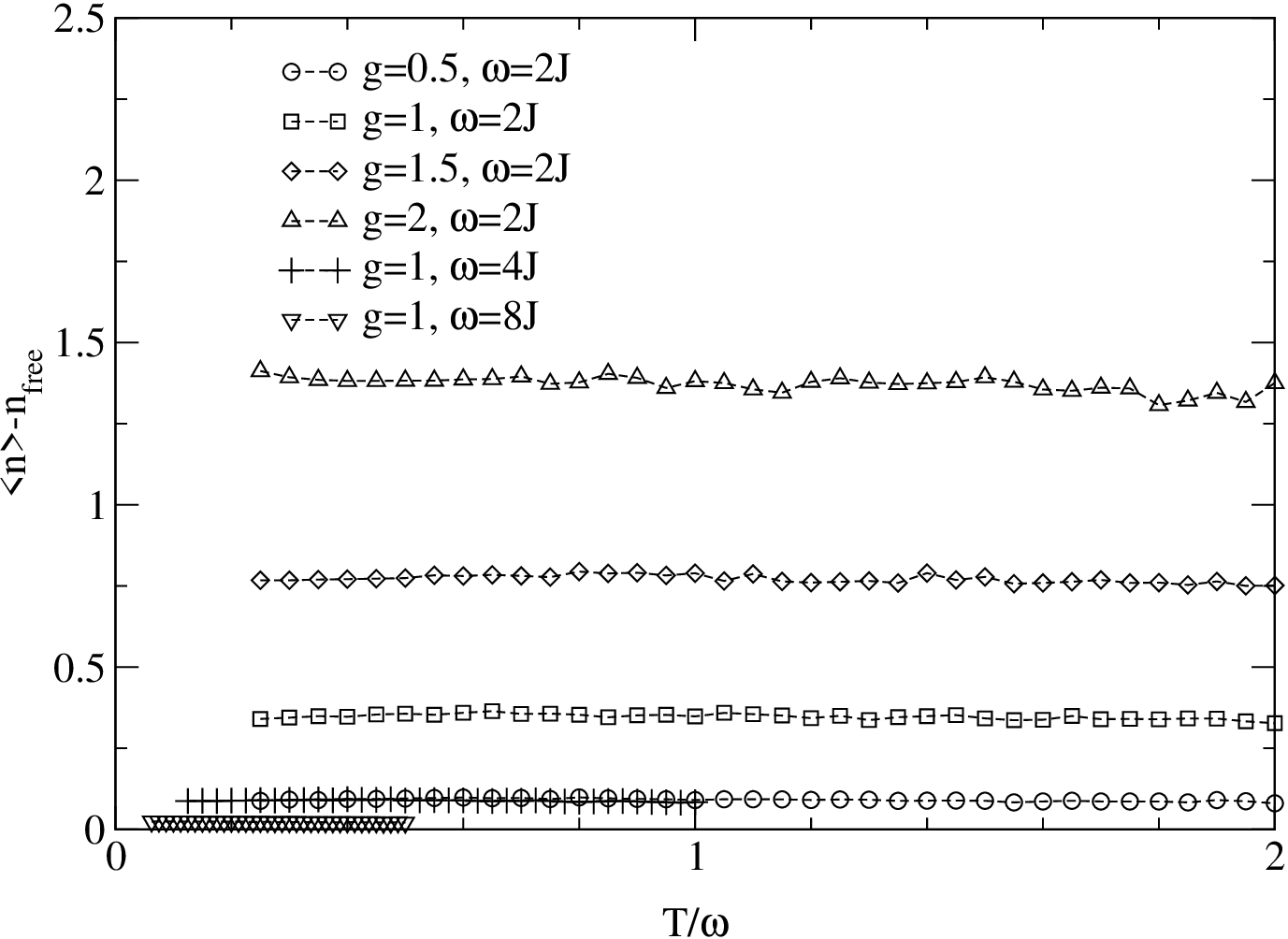}
\end{center}
\caption{Differences $\langle n \rangle-n_\text{free}$ vs.~$T/\omega$.\label{ngvgla}}
\end{figure}

As expected, we find no difference in the mean occupation numbers $\langle n \rangle$ and $\langle m \rangle$, and therefore restrict the following discussion to the behaviour of $\langle n \rangle$. Figure \ref{nga} shows the Monte Carlo results for $\langle n \rangle$ for different values of $g$ and $\omega$ in a plot vs.~$T/\omega$. The data are compared to the Bose distribution for free Einstein phonons
\begin{equation}\label{nfree}
n_\text{free}(T)=\frac{1}{e^{\omega/T}-1},
\end{equation}
which is also shown in \tfig\ref{nga}. Again we find a striking similarity to the 1D bond coupling model. In $d=1$ it has been found that the mean phonon occupation numbers obey the relation\cite{kuehne99} 
\begin{equation}\label{nmedium}
\langle n \rangle (T)= n_0 + n_\text{free}(T)
\end{equation}
with a temperature-independent constant $n_0$. As a good approximation, this relation is valid in a temperature range $0.5J\leq T\leq 3J$ in the 2D case as well. This can be seen most clearly in \tfig\ref{ngvgla}, which shows the same results as \tfig\ref{nga} with relation \eqref{nfree} subtracted from the Monte Carlo data.
\begin{figure}[b!]
\begin{center}
  \includegraphics[angle=270,width=\linewidth]{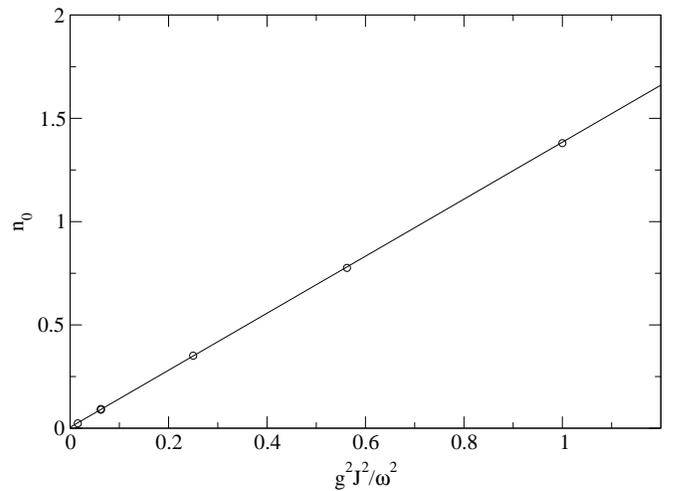}
\end{center}
\caption{Over the temperature range $0.5J\leq T\leq 3J$ averaged values $n_0$ vs.~$g^2J^2/\omega^2$. The solid line shows the result from linear regression.\label{ausgg}}
\end{figure}

In order to derive an expression for $n_0$, we averaged the differences $\langle n \rangle-n_\text{free}$ over the temperature range \mbox{$0.5J\leq T\leq 3J$} and plotted these values vs.~$g^2 J^2/\omega^2$ (see \tfig\ref{ausgg}). By applying linear regression we find that in $d=2$ the shift obeys the relation
\begin{equation}\label{n0}
n_0\approx (1.375 \pm 0.003)\left(\frac{gJ}{\omega}\right)^2.
\end{equation}
Thus the only difference between the relations for the mean phonon occupation numbers in the 1D and 2D case is given by the numerical prefactor in \eqref{n0}, the value being 1.375 in $d=2$ and 2 in $d=1$.\cite{kuehne99}  

We close this section with a technical remark concerning our choice for the cutoff for the phonon occupation numbers. As can be seen in \tfig\ref{nga}, in the whole temperature range the measured mean occupation numbers are more than an order of magnitude smaller than the cutoff 40. In retrospect our choice therefore is justified. With \eqref{nmedium} and \eqref{n0} we have also found an expression that might be important for other numerical methods (\eg exact diagonalization) which depend crucially on a (low) cutoff in the phonon numbers. 

\subsection{Specific heat}\label{kapaz}
\begin{figure}[b!]
\begin{center}
  \includegraphics[width=\linewidth]{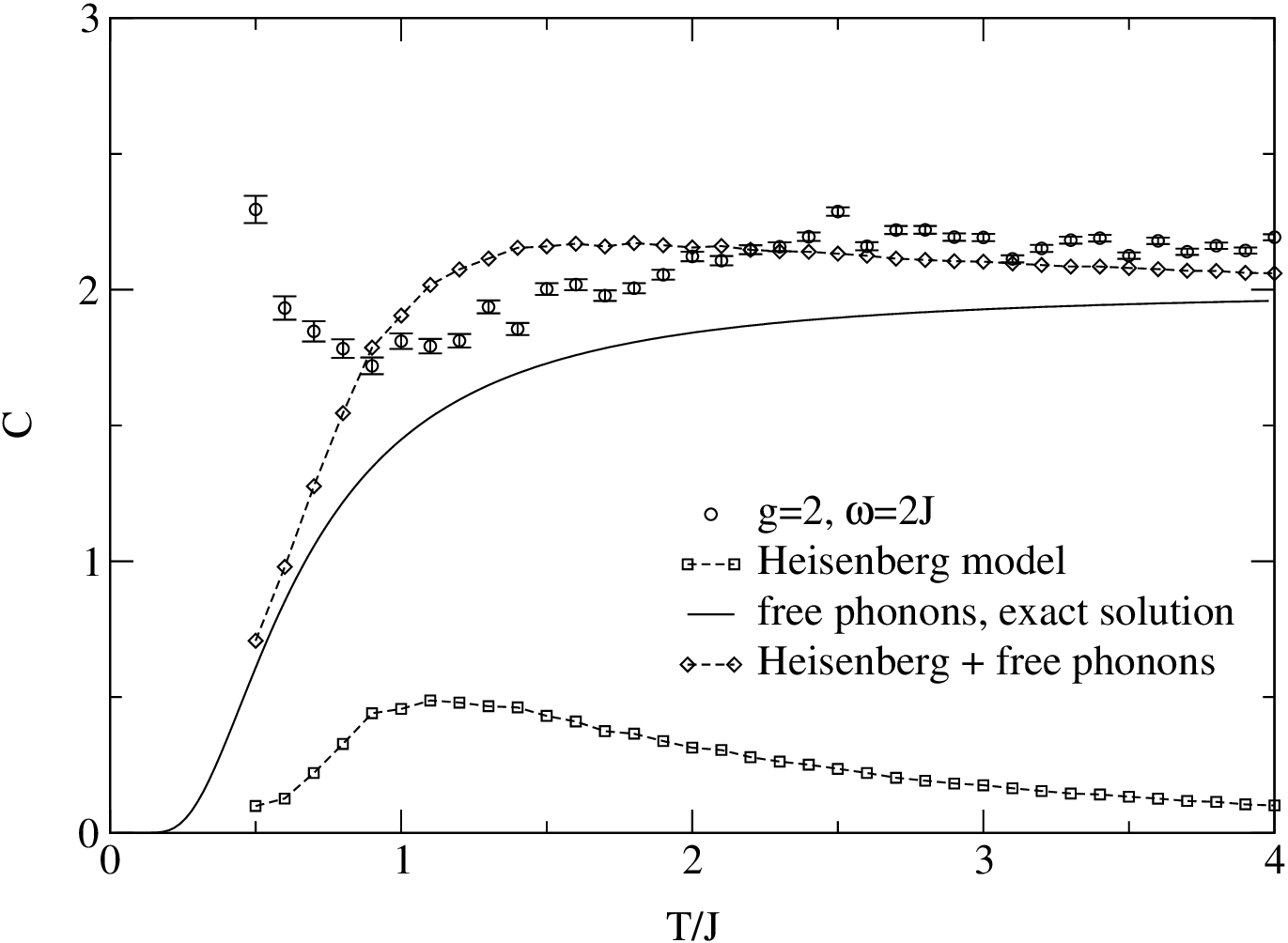}
\end{center}
\caption{Specific heat plotted vs.~temperature for $g=2$ and $\omega=2J$. The solid line shows result \eqref{heatcap} for free phonons of the same frequency. Plotted are also Monte Carlo results for the 2D Heisenberg model and the sum of the free phonon and Heisenberg results. The number of spin updates is 5$\times10^5$ for the system with spin phonon coupling and $10^5$ in the Heisenberg case.\label{csummea}}
\end{figure}

Another important thermodynamic quantity is the specific heat per site $C$. Although in principle this observable is directly accessible in the experiment, it is dominated by lattice vibrations, making it difficult to extract its magnetic part. Even for a simple model as given by Hamiltonian \eqref{ham}, we find this behaviour confirmed. Figure \ref{csummea} shows Monte Carlo data for the specific heat in a system with $g=2$ and $\omega=2J$ and the exact result
\begin{equation}\label{heatcap}
C_\text{free}(T)=2\left(\frac{\omega}{T}\right)^2 \frac{e^{\omega/T}}{(e^{\omega/T}-1)^2}
\end{equation}
for free phonons of the same frequency (the factor two accounts for two phonons per lattice site). There is only a small difference in the overall height of $C$ and $C_\text{free}$. At high temperatures, both curves approach the same constant value, yielding the Dulong-Petit rule.

The spin phonon coupling influences the specific heat significantly though. As can be seen in \tfig\ref{csummea} as well, the curve for the system with spin phonon coupling differs significantly from the sum of Monte Carlo results for the 2D Heisenberg model and the contribution \eqref{heatcap} for free Einstein phonons. Note that both the strong fluctuations and the divergency of the data for $T\rightarrow 0$ are due to difficulties in evaluating the specific heat within Monte Carlo procedures as discussed in \tref\onlinecite{fye87}.

\section{Spin correlation function and ground state properties}\label{2dgs}

We now turn our attention towards ground state properties of model \eqref{ham}. In principle the Monte Carlo method is only applicable at finite temperatures. By analyzing the behaviour of the spin correlation function at low temperatures, however, it is possible to make statements about system properties at $T=0$.  
\begin{figure}
\begin{center}
  \includegraphics[width=\linewidth]{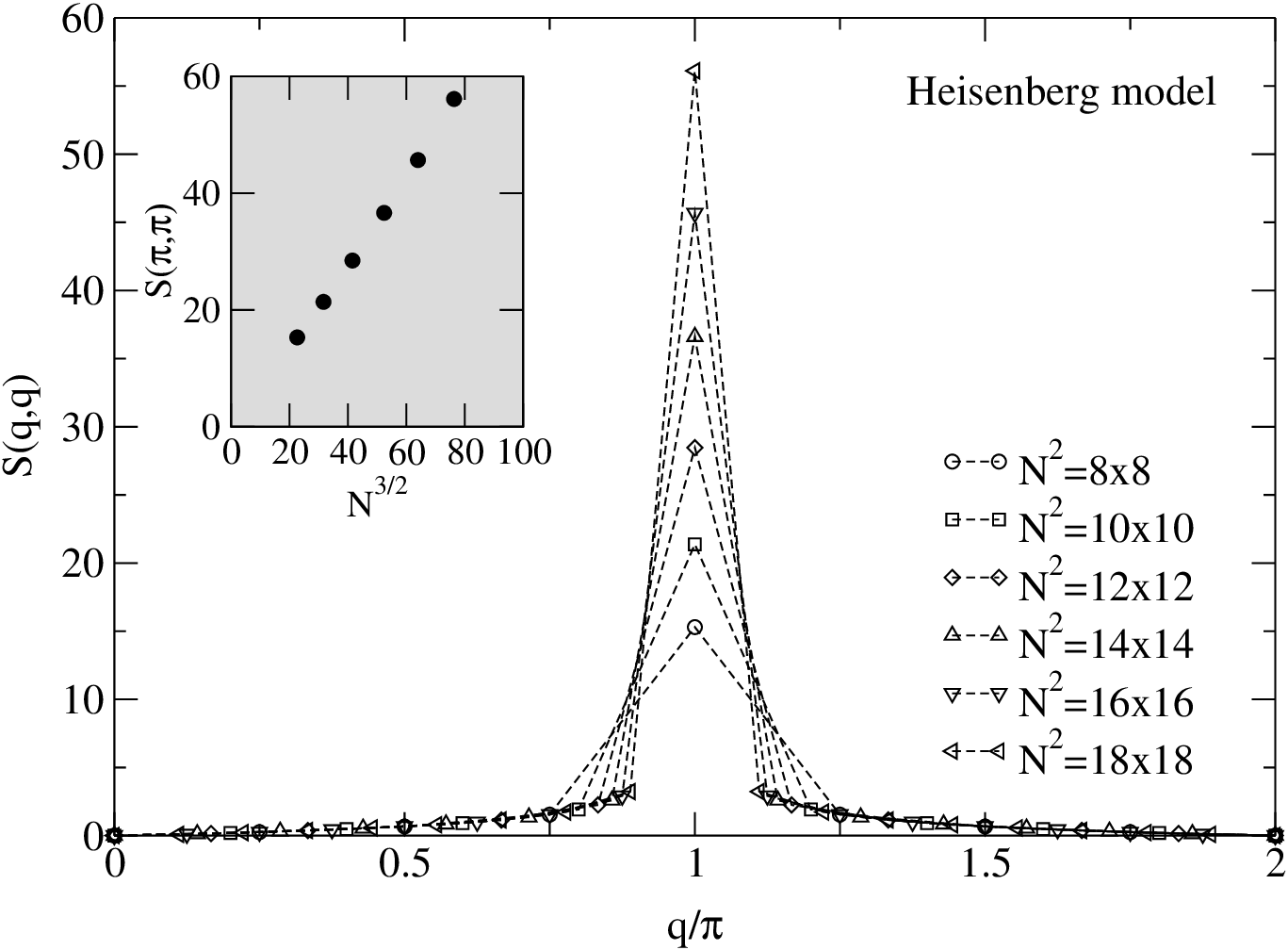}
  \includegraphics[width=\linewidth]{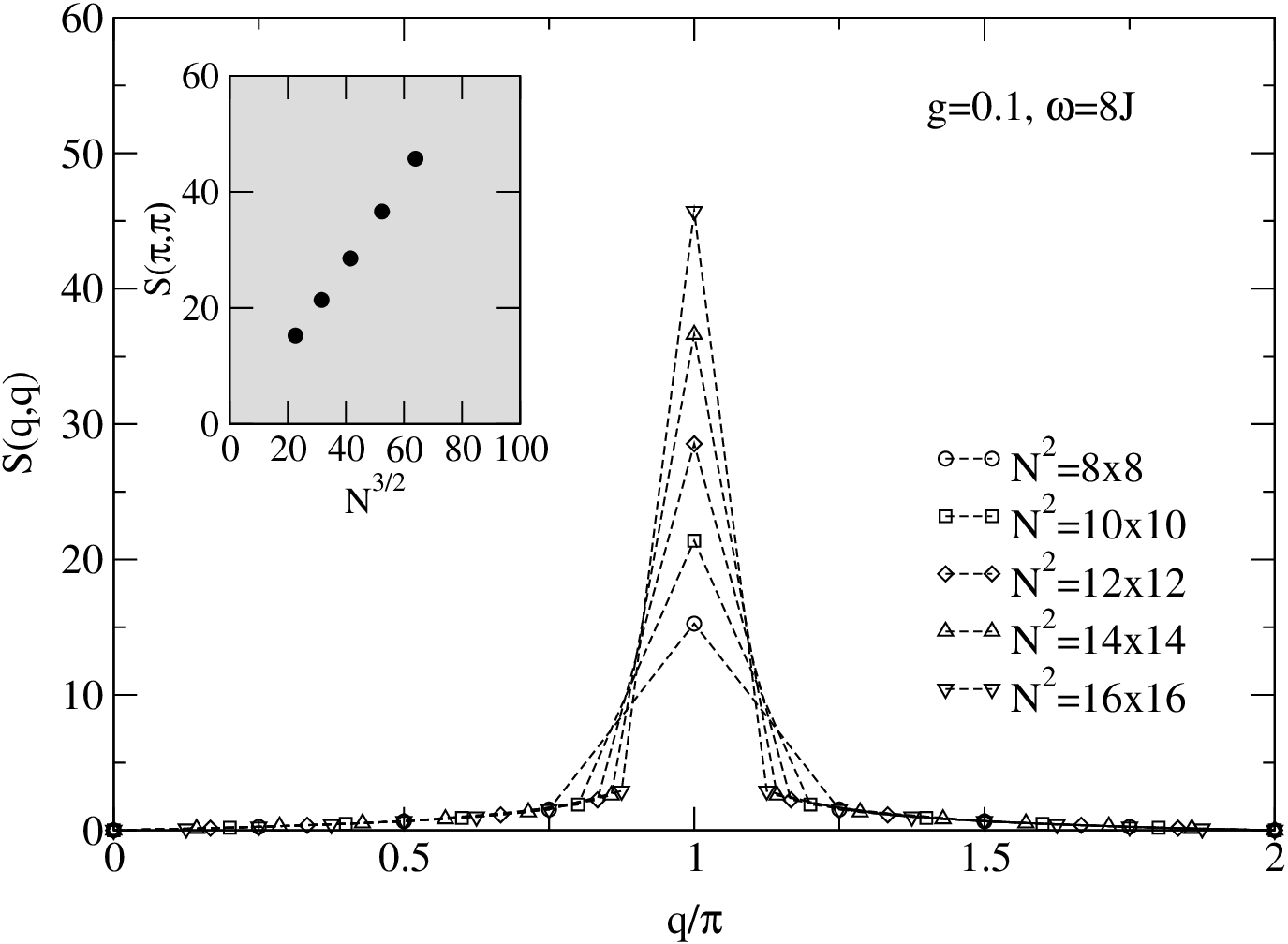}
  \includegraphics[width=\linewidth]{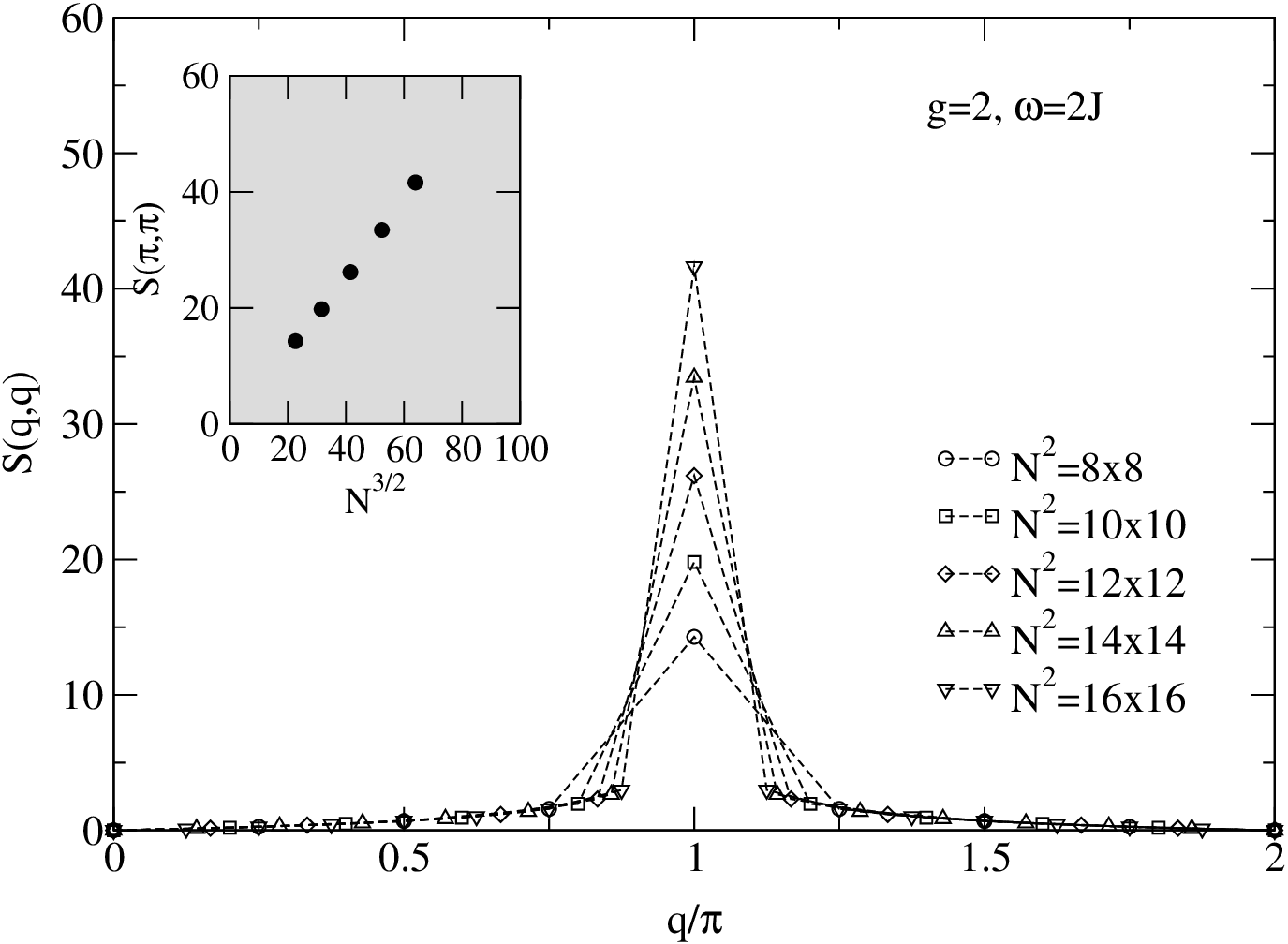}
\end{center}
\caption{Static structure factor $S(q,q)$ of spin correlations as a function of momentum $q$ for a diagonal cut through the first Brillouin zone at $T=0.1J'$ for different system sizes. Shown are results for the 2D Heisenberg model (top) and for model \eqref{ham} with $g=0.1$, $\omega=8J$ (middle) and $g=2$, $\omega=2J$ (bottom), respectively. In all plots the inset shows the height of the maximum at $(\pi,\pi)$ as a function of $N^{\frac{3}{2}}$.\label{st2d}}
\end{figure}

The argument is as follows. Suggest that we choose the coupling constants in \eqref{ham} such that the system is Heisenberg-like, showing long range N\'eel order in the ground state. Then for reasons of universality we expect that the spin correlation function
\begin{equation}\label{spinspin}
G(\vec{d}\,)=\frac{1}{N^2}\sum_{\vec{r}}\langle \vec{\sigma}_{\vec{r}\,}\vec{\sigma}_{\vec{r}+\vec{d}}\,\rangle
\end{equation}
obeys the result known for the 2D Heisenberg model\cite{chakra89,ding90,barnes91}
\begin{equation}\label{heisenberg}
G(\vec{d}\,)\sim (-1)^{d_1+d_2}\; |\vec{d}\,|^{-\lambda}\; e^{-|\vec{d}\,|/\xi(T)},
\end{equation}
with the algebraic exponent $\lambda$ close\cite{ding90} to the classical Ornstein-Zernike value of $\frac{1}{2}$. Here $\xi(T)$ is the spin correlation length which can be interpreted as the mean size of domains with antiferromagnetic order. At $T=0$ these domains get macroscopic, because for $T\rightarrow 0$ the correlation length diverges exponentially.\cite{chakra89,ding90,barnes91,beard98} Assuming $\lambda=\frac{1}{2}$ in \eqref{heisenberg}, this means that the static structure factor  
\begin{equation}\label{struktur}  
S(\vec{q}\,)=\sum_{\vec{d}}e^{i\vec{q}\vec{d}}G(\vec{d}\,)
\end{equation}
diverges for momentum\footnote{In the following we set $a\equiv 1$ for the lattice constant.} $\vec{q}=(\pi,\pi)$ with the linear system size like $N^{\frac{3}{2}}$. As long as the correlation length in the infinite system stays significantly larger than the system sizes considered, this behaviour should be visible at low temperatures.

We first illustrate this in case of the 2D Heisenberg model ($g=\omega=0$ in \eqref{ham}) at $T=0.1J$, taking $10^5$ spin updates and choosing a Trotter number of $M=160$ for the evaluation of spin correlations. At this temperature, the correlation length in the infinite system is of the order of $10^9$ lattice spacings.\cite{chakra89,ding90,barnes91,beard98} As can be seen in the upper plot of \tfig\ref{st2d}, the static structure factor shows a pronounced maximum at $\vec{q}=(\pi,\pi)$, and the peak height roughly scales with the system size like $N^\frac{3}{2}$. Note that deviations from the $N^\frac{3}{2}$-dependence might indicate that in the quantum system $\lambda$ differs slightly from the value of $\frac{1}{2}$ chosen above.  
 
We now return to model \eqref{ham} with spin phonon coupling and discuss our results for two different choices of coupling constants. To compare our results for different values of $g$ and $\omega$ we drop the unphysical terms $-Jg/2(a_{ij}^\dagger + a_{ij}),-Jg/2(b_{ij}^\dagger + b_{ij})$ in \eqref{ham}. We therefore measure the temperature in units of the rescaled magnetic coupling $J'$ of the effective Hamiltonian \eqref{hameff}. We calculated spin correlations at $T=0.1J'$, taking $5\times 10^5$ spin updates and $M=160$ in our calculations. First, we consider a system with a small value for $\frac{gJ}{\omega}$ ($g=0.1$, $\omega=8J$) where no dimerization is expected. Again the static structure factor shows a pronounced peak for $\vec{q}=(\pi,\pi)$, and the peak height scales with the system size like $N^\frac{3}{2}$ (middle of \tfig\ref{st2d}), indicating Heisenberg-like behaviour as anticipated. 

For the second system we choose $g=2$ and $\omega=2J$. In the 1D case this choice corresponds to a system which strongly dimerizes in the ground state.\cite{raas02} Even here we find a pronounced peak of $S$ for $(\pi,\pi)$ (bottom of \tfig\ref{st2d}), the maximum height scaling like $N^\frac{3}{2}$. The interpretation is that even in the regime of large values for $\frac{gJ}{\omega}$ the system shows antiferromagnetic order in the ground state. This is a striking difference to the 1D bond coupling model. Compared to the case with $g=0.1$ and $\omega=8J$, however, we find that the spin phonon coupling counteracts the tendency of the system to order antiferromagnetically in the ground state. As can be seen in \tfig\ref{st2d}, for fixed system size the peak heights $S(\pi,\pi)$ in the case of strong spin phonon coupling are slightly diminished as compared to the weak coupling regime.
 
Our results can be confirmed by a direct analysis of the temperature dependence of the spin correlation length. For both systems and at various temperatures we extracted finite system correlation lengths $\xi_N$ by fitting the function
\begin{equation}\label{symmetrie}
f(\vec{d}\,)=a (-1)^{d_1+d_2}\left(\frac{e^{-|\vec{d}\,|/\xi}}{\sqrt{|\vec{d}\,|}}+\frac{e^{-(N-|\vec{d}\,|)/\xi}}{\sqrt{N-|\vec{d}\,|}}\right)
\end{equation}
with two free parameters $a,\xi$ to our data for system sizes \mbox{$N=10,12,14,20,24$}. For $g=0.1$ and $\omega=8J$, we selected values $M=120$ for temperatures $0.5J\leq T\leq 0.9J$ and $M=80$ for $T\geq J$, taking $10^5$ spin updates (for $N=24$ we chose $M=120$ for all temperatures). For $g=2$ and $\omega=2J$ the choice was $M=160$ for $1.5J\leq T\leq 1.9J$, $M=120$ for $2J\leq T\leq 4J$ and $M=80$ for $T\geq 5J$, again averaged over $10^5$ Monte Carlo sweeps (for $N=20$ we took $1.5\times10^5$, for $N=24$ and $1.5J\leq T\leq 1.9J$ we took $2\times10^5$ spin updates). The values for the Trotter number are sufficiently large to avoid finite size effects in Trotter direction.   

\begin{figure}[t!]
\begin{center}
  \includegraphics[width=\linewidth]{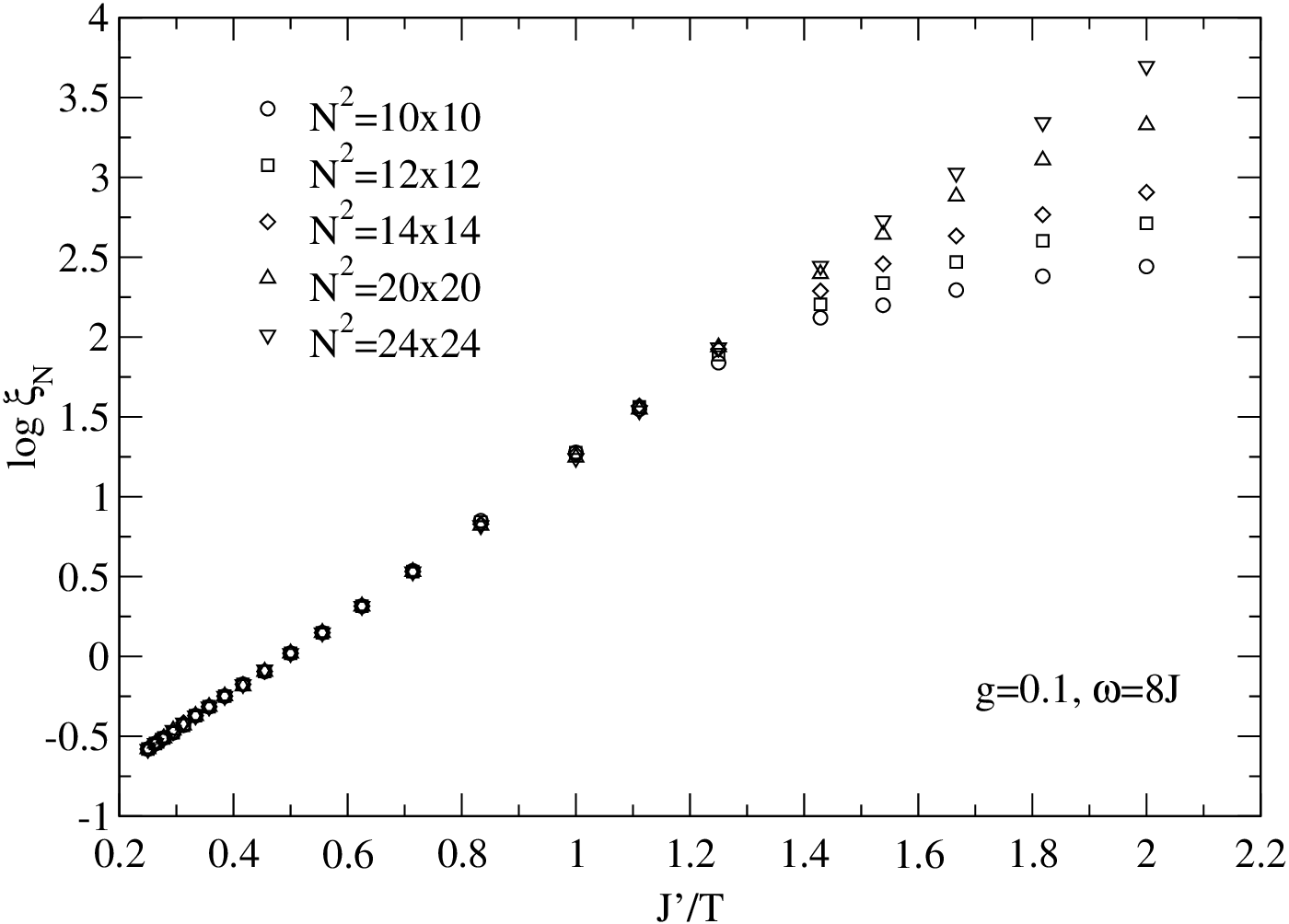}
  \includegraphics[width=\linewidth]{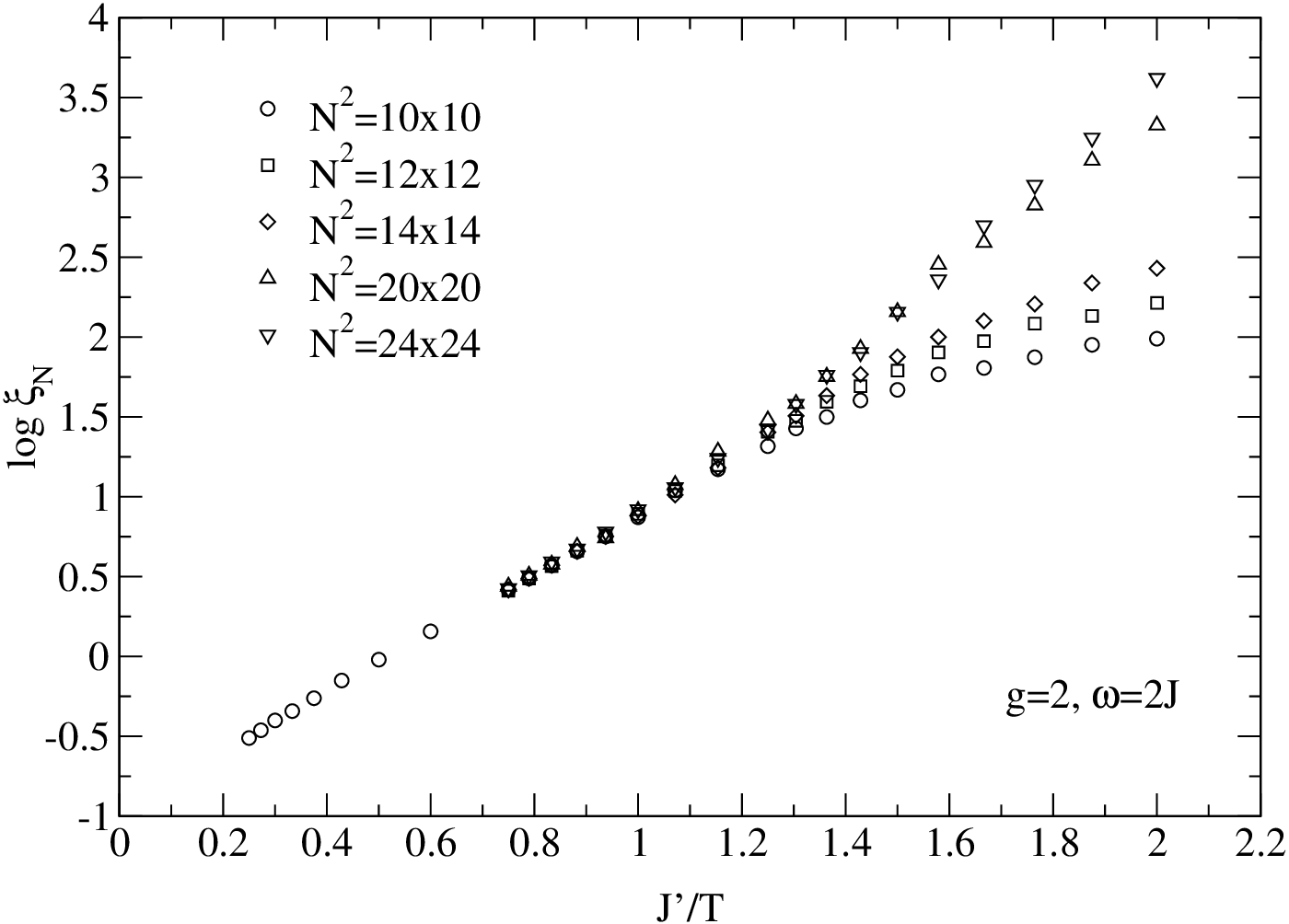}
\end{center}
\caption{Natural logarithm of finite system correlation lengths $\xi_N$ vs.~inverse temperature for five different system sizes for \mbox{$g=0.1$}, $\omega=8J$ (top) and $g=2$, $\omega=2J$ (bottom), respectively. Note that all temperatures are given in units of $J'$.\label{xi2d}}
\end{figure}

As has been said above, in case of the Heisenberg model in leading order $\xi$ behaves like $e^\frac{J}{T}$ at low temperatures.\cite{chakra89,ding90,barnes91,beard98} In the upper panel of \tfig\ref{xi2d} the natural logarithm of $\xi_N$ is plotted vs.~$\frac{1}{T}$ for $g=0.1$ and $\omega=8J$. At high temperatures, no dependence on the system size is visible, and as expected we find the same behaviour as in the Heisenberg-model. At low temperatures finite size effects become important and the curves branch off from the asymptotic linear behaviour. In the case of strong spin phonon coupling the graph shows very similar features (bottom of \tfig\ref{xi2d}). We therefore find Heisenberg-like behaviour of $\xi$ at finite temperatures also in the regime of strong spin phonon coupling. The main difference between the two cases is that at the same effective temperature the correlation length for $g=2$, $\omega=2J$ is significantly smaller than for $g=0.1$, $\omega=8J$. The analysis of $\xi$ therefore also implies that a strong spin phonon coupling weakens antiferromagnetic order. Both observations are consistent with the conclusions drawn from the analysis of $S(\vec{q}\,)$. Note that in both plots of \tfig\ref{xi2d} the temperatures are given in units of $J'$. 

\begin{figure}
\begin{center}
  \includegraphics[width=\linewidth]{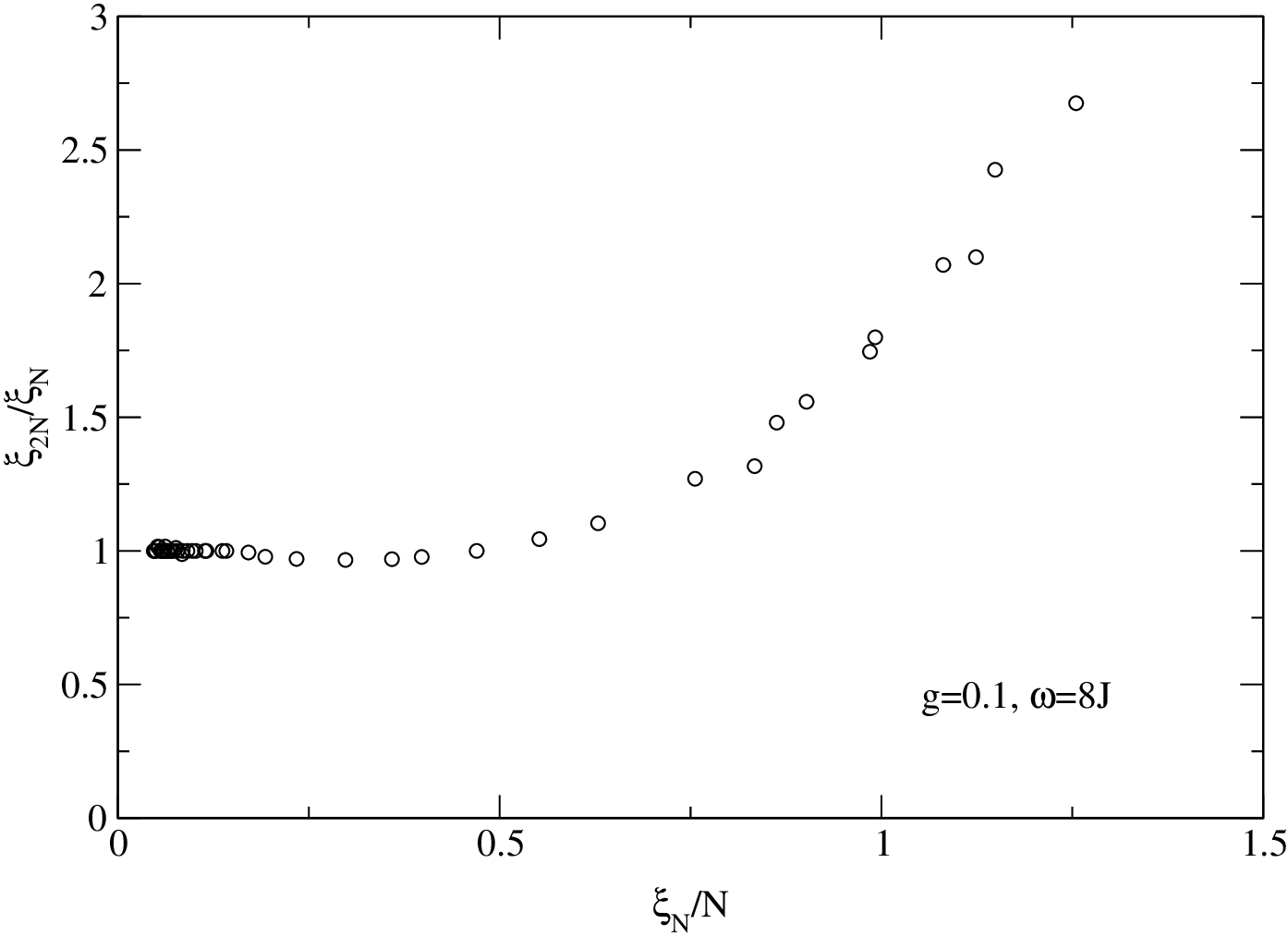}
  \includegraphics[width=\linewidth]{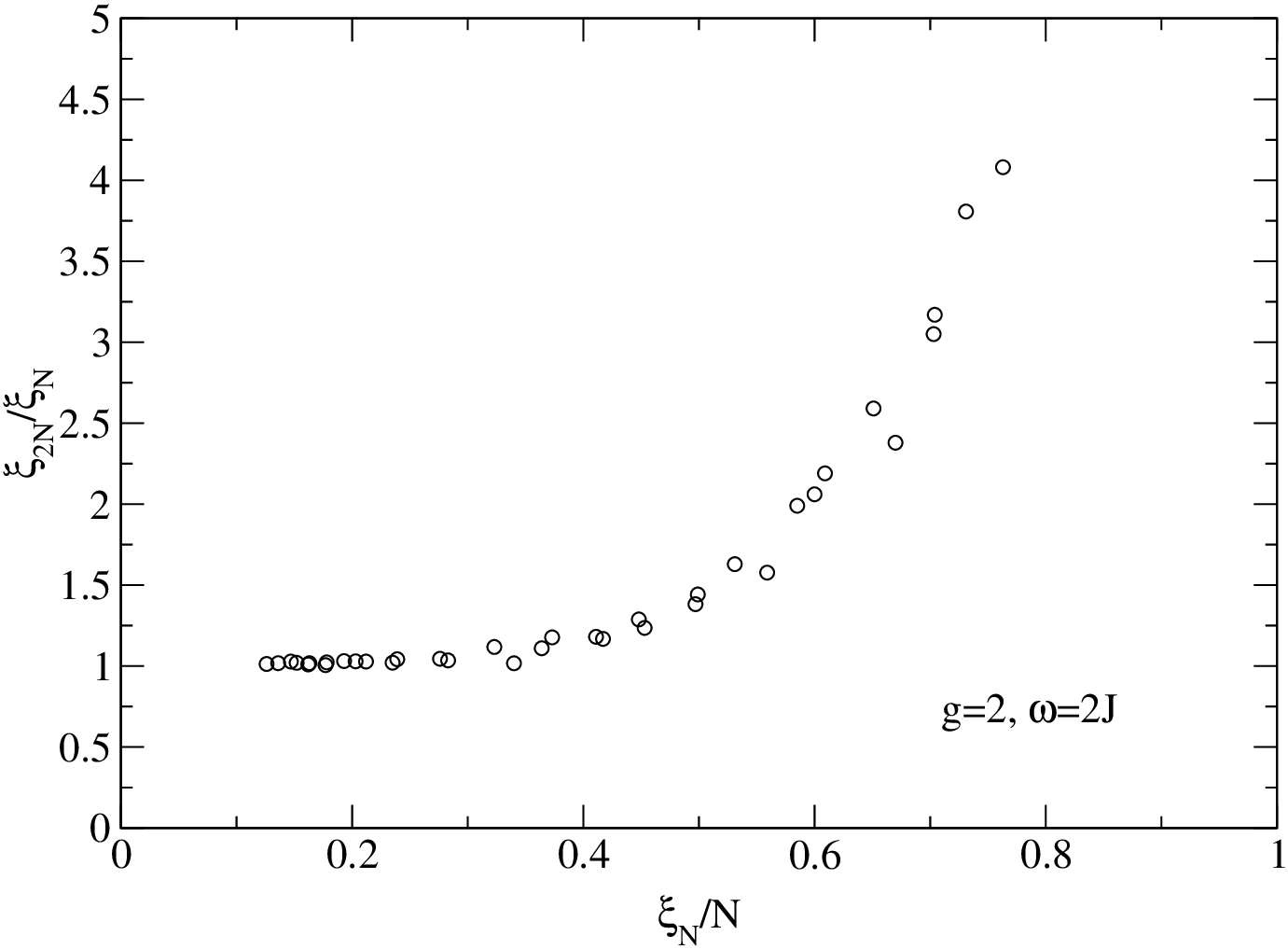}
\end{center}
\caption{Test of the scaling prediction \eqref{hypo} for $g=0.1$, \mbox{$\omega=8J$} (top) and $g=2$, $\omega=2J$ (bottom).\label{scaling2d}}
\end{figure}
By means of scaling arguments our analysis can be extended to make direct statements about ground state properties. Suppose the system shows long range N\'eel order in the ground state. In terms of the renormalization group this means that there is a critical fixed point at $T=0$ which controls the system properties at low temperatures. In this case a finite size scaling ansatz\cite{cara95,cara95a,beard98}
\begin{equation}\label{hypo}
\frac{\xi_{2N}(T)}{\xi_{N}(T)}=F\left(\frac{\xi_N(T)}{N}\right)
\end{equation}
holds, where $F$ is a universal scaling function. With the data from \tfig\ref{xi2d} it is possible to test the scaling prediction \eqref{hypo}. Plotting $\frac{\xi_{2N}}{\xi_{N}}$ vs.~$\frac{\xi_N}{N}$ with $N=10,12$ we find that both for $g=0.1$, $\omega=8J$ (top of \tfig\ref{scaling2d}) and $g=2$, $\omega=2J$ (bottom of \tfig\ref{scaling2d}) the data lie on one curve. The shape of the scaling function $F$ in \eqref{hypo}, however, depends on the choice of the coupling constants. The interpretation is that in the weak and in the strong coupling regime model \eqref{ham} shows long range N\'eel order in the ground state, strongly confirming the conclusions drawn from our analysis of the static structure factor at low temperatures.
  
\begin{figure}
\begin{center}
  \includegraphics[width=\linewidth]{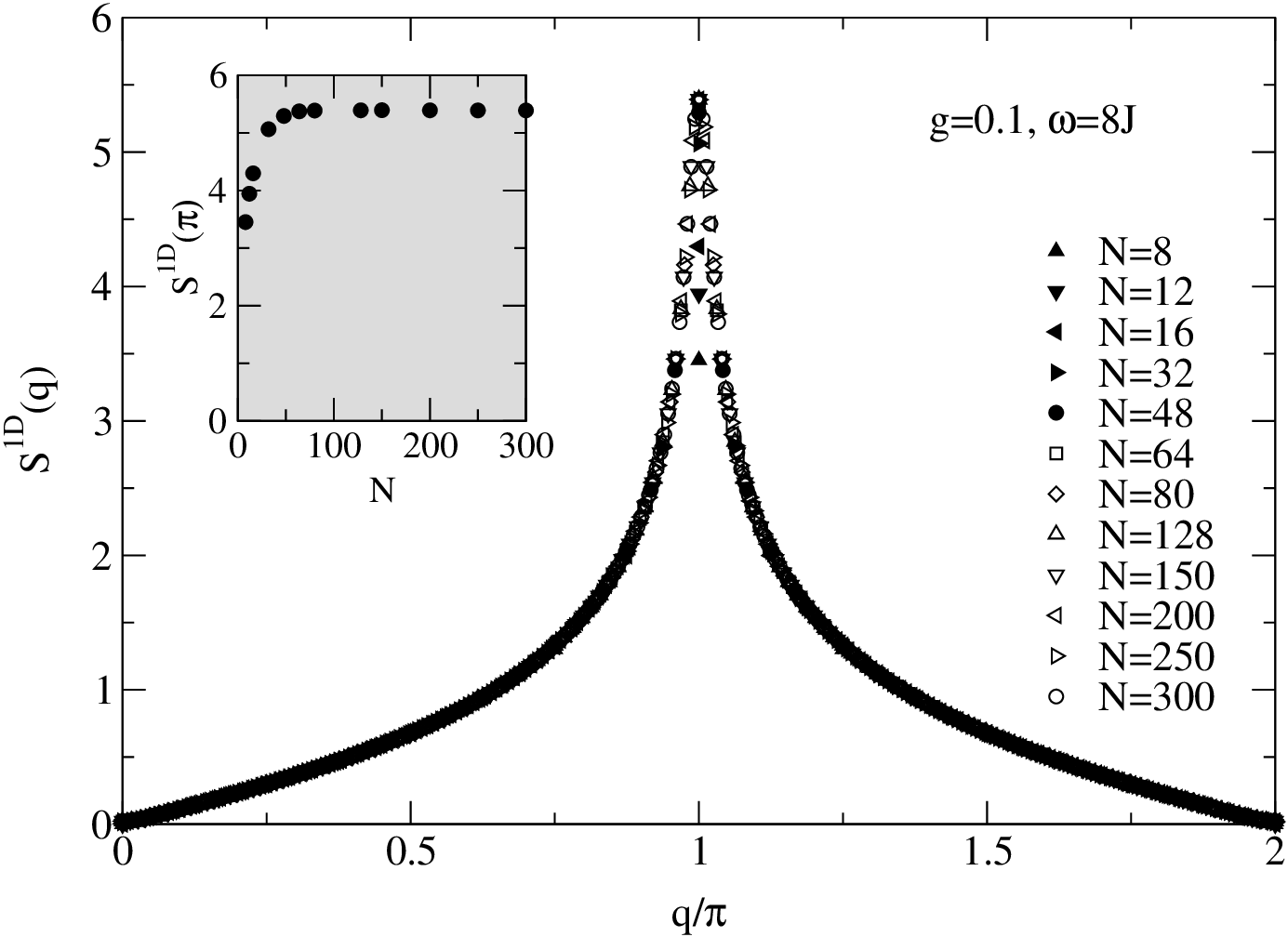}
  \includegraphics[width=\linewidth]{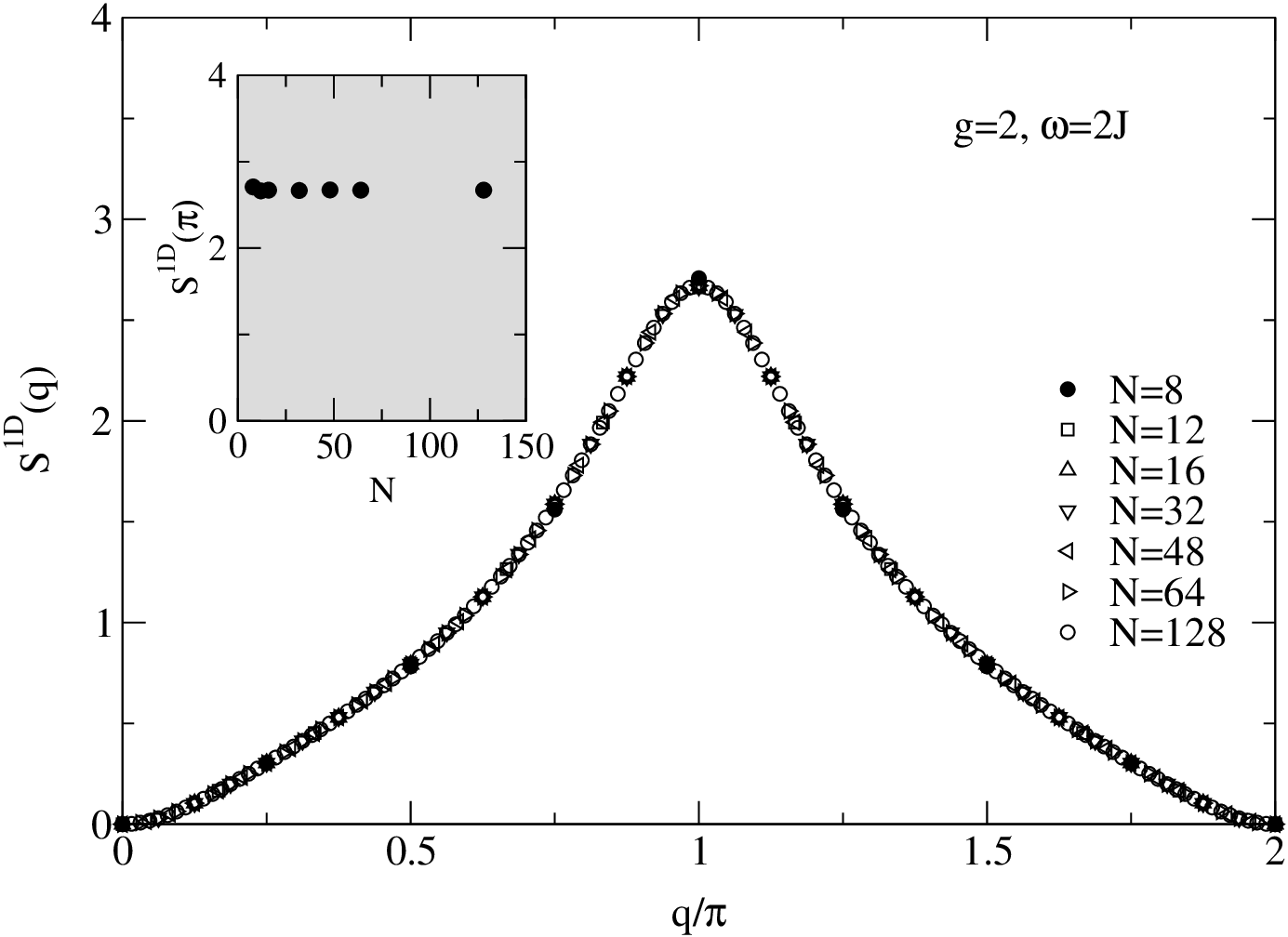}
\end{center}
\caption{Static structure factors $S^\text{1D}(q)$ of spin correlations vs.~momentum $q$ in the 1D case for different system sizes and temperature $T=0.1J'$. In the upper graph the coupling constants are $g=0.1$ and $\omega=8J$, in the lower graph $g=2$ and $\omega=2J$. In both plots the inset shows the height of the maximum $S^\text{1D}(\pi)$ as a function of $N$.\label{st1d}}
\end{figure}
It is instructive to compare these results to the low temperature behaviour of spin correlations in \mbox{$d=1$}. Our argumentation is completely analogous to the 2D case. As has been discussed in \tchap\ref{intro}, the 1D model shows a quantum phase transition between a N\'eel ordered and a dimerized phase.\cite{weisse99b,kuehne99,raas02} Though only the approximate shape of the phase separation line in coupling constant space has been determined, it is known that for small values $\frac{gJ}{\omega}$ the system is Heisenberg-like, showing quasi long range N\'eel order in the ground state. This means that the 1D correlation function 
\begin{equation}\label{spin1D}
G^\text{1D}(d)=\frac{1}{N}\sum_{i=1}^{N} \langle \vec{\sigma}_i \vec{\sigma}_{i+d}\rangle
\end{equation}
decays exponentially at finite temperatures, with a rate given by the correlation length $\xi^\text{1D}\propto \frac{1}{T}$.\cite{takada86,suzuki90,nomura91,suzuki92,kluemper93}
At $T=0$, there is a crossover to an algebraic decay\cite{luther75,fogedby78,korepin84,bogo86} 
\begin{equation}\label{gz1D}
G^\text{1D}(d)\sim \frac{(-1)^d}{d},
\end{equation}
and at $T=0$ the static structure factor
\begin{equation}
S^\text{1D}(q)=\sum_{d=1}^N e^{iqd} G^\text{1D}(d)
\end{equation}
diverges for momentum $q=\pi$ like the $N$th partial sum of the harmonic series with the system size. As in $d=2$, signs of this divergence should be visible at low temperatures. 

For large values $\frac{gJ}{\omega}$, on the other hand, the chain dimerizes. This means that long range dimer order is established in the ground state. Therefore the spin correlation length $\xi^\text{1D}$ stays finite at $T=0$, and we expect no dependence of $S^\text{1D}$ on $N$ in the low temperature regime.

In \tfig\ref{st1d} the static structure factors $S^\text{1D}$ for two systems with the same choice of coupling constants as in $d=2$ are plotted. As in $d=2$, we measure the temperature in units of the rescaled magnetic coupling $J'$ of the 1D counterpart of \eqref{hameff}. We selected $T=0.1J'$ and $M=160$, executing $5\times 10^5$ spin updates. In both systems $S^\text{1D}$ shows a maximum at $q=\pi$. In the Heisenberg-like system the maximum is more pronounced though, and for small system sizes the peak height depends on $N$. For larger system sizes, however, such a behaviour is not visible. This leads to the conclusion that in contrast to the 2D case the correlation length in the infinite system is not larger than the system sizes in consideration. In the system with dimerization in the ground state we find the expected behaviour: The values $S^\text{1D}(\pi)
$ do not depend on the system size, indicating that $\xi^\text{1D}$ is very small.

\begin{figure}
\begin{center}
  \includegraphics[width=\linewidth]{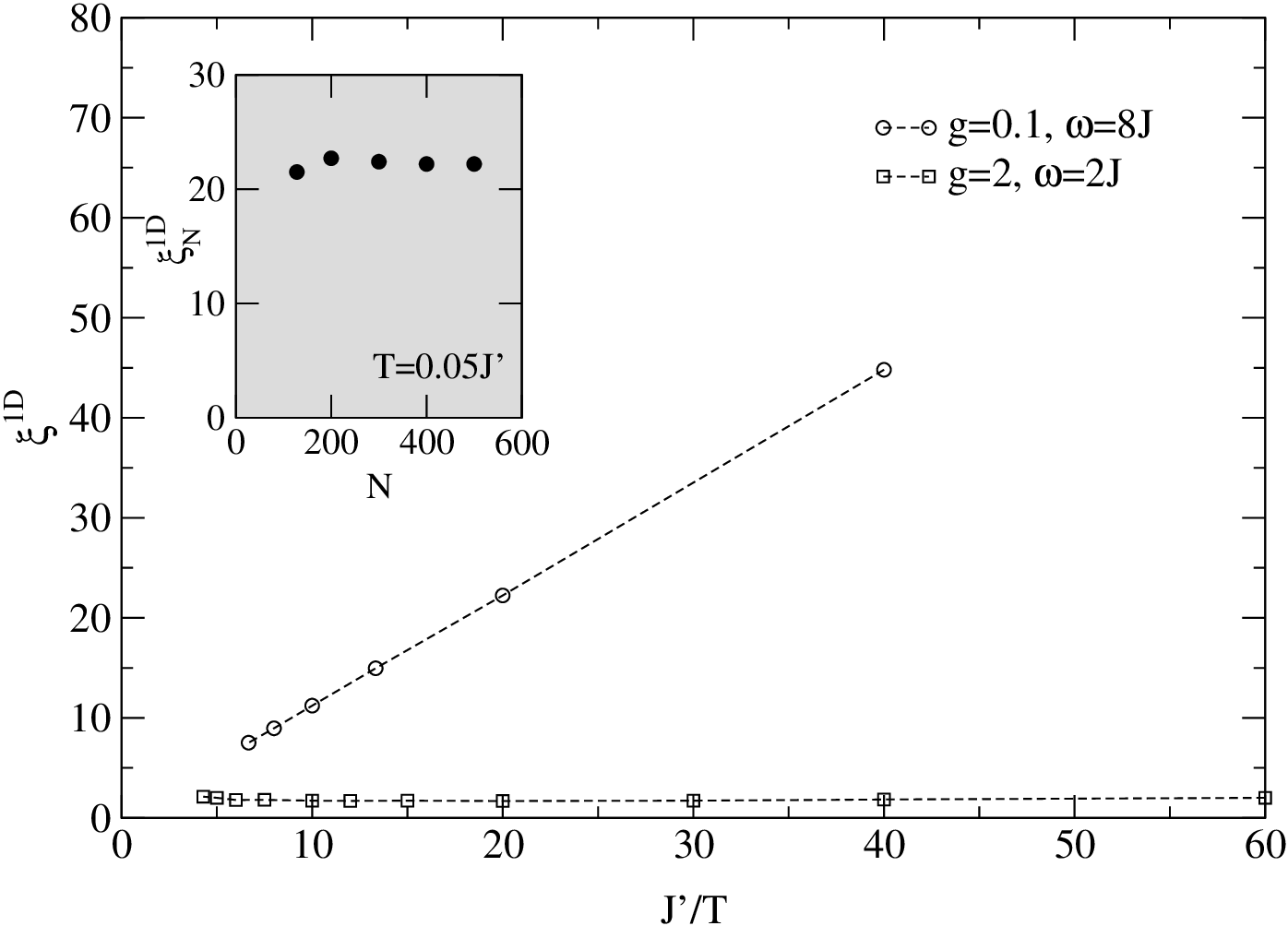}
\end{center}
\caption{Correlation lengths $\xi^\text{1D}$ vs.~inverse temperature in the 1D case. Note that the temperatures are given in units of $J'$. The inset shows the finite size behaviour of $\xi^\text{1D}_N$ for $g=0.1$, $\omega=8J$ at $T=0.05J'$.\label{xi1d}}
\end{figure}

By analyzing the temperature dependence of $\xi^\text{1D}$ it is possible to distinguish more clearly between the two regimes. Similar to $d=2$ we extracted correlation lengths $\xi^\text{1D}$ by fitting an exponential decay with two free parameters to our data. For $g=0.1$ and $\omega=8J$, we executed $5\times 10^5$ spin updates and selected $M=160$, which is large enough to avoid effects by the finite Trotter number. Furthermore, the system sizes were chosen that large that finite size effects are negligible ($N=500$ for $T=0.025J,0.05J$, $N=400$ for $T=0.075J,0.125J$ and $N=300$ for $T=0.1J,0.15J$). This can be seen in the inset of \tfig\ref{xi1d}, where $\xi^\text{1D}_N$ is plotted vs.~$N$ at $T=0.05J\approx 0.05J'$. For $g=2$ and $\omega=2J$ also $5\times 10^5$ spin updates were made. The correlation lengths are that small that a chain length of $N=200$ is sufficient to make statements about the thermodynamic limit. The effect of the finite Trotter number is more important in this system. For $M=400$ at the lowest temperatures, however, the effect is smaller than the error which enters our analysis during the fitting procedure (we chose $M=400$ for $T=0.05J,0.075J,0.1J$, $M=360$ for $T=0.15J$ and $M=160$ for $T\geq 0.2J$).

Figure \ref{xi1d} shows the correlation lengths in a plot vs.~the inverse temperature in units of the rescaled effective coupling $J'$. In both systems we find the expected behaviour: In the Heisenberg-like system the correlation length grows linearly with the inverse temperature, while in the dimerized system $\xi^\text{1D}$ takes small values and shows no such dependence in the temperature range shown. 

We close this section with a final remark concerning the 2D model \eqref{ham}. The results from this section need not mean that the model shows no lattice distortion. In both the statically dimerized stair and plaquette models e.~g. a phase with coexisting dimerization and long range antiferromagnetic order is known.\cite{sirker02} For model \eqref{ham} it is therefore conceivable that small lattice distortions appear which cannot be detected by analyzing spin correlations at low temperatures. Even a finite temperature phase transition seems possible, because due to the Mermin-Wagner theorem\cite{mermin66} a breaking of the discrete lattice symmetry at finite temperatures cannot be excluded in a 2D system. However, we find no hints on a finite temperature phase transition in the behaviour of thermodynamic properties in the temperature range discussed in \tchap\ref{2dtd}. Further investigations of the model seem appropriate to clarify whether dimerization appears and which dimerization pattern is realized. 

\section{Summary}\label{sum}

By combining loop updates for spin and cluster updates for phononic degrees of freedom we have developed a quantum Monte Carlo algorithm to study the properties of the 2D antiferromagnetic Heisenberg model coupled to bond phonons. 

As thermodynamic quantities are concerned, we studied the susceptibility, mean phonon occupation numbers and specific heat in the temperature range \mbox{$0.5J\leq T\leq 4J$}. The properties of the model at finite temperatures are similar to the 1D case.

For temperatures $0.5J\leq T\leq 3J$, we derived an expression for the mean phonon occupation numbers which is of practical value for further investigations of the model.

We investigated the temperature dependence of the spin correlation length for two choices of coupling constants. Our analysis indicates that the model shows long range N\'eel order in the ground state both in the regime of weak and strong spin phonon coupling.

\begin{acknowledgments}
The authors would like to thank M.\@ Braden, A.\@ Kl\"umper, E.\@ M\"uller-Hartmann, C.\@ Raas, W.\@ Weber and R.\@ Werner for valuable discussions. We acknowledge funding by the DFG in the Sonderforschungsbereich 608.
\end{acknowledgments}

\end{document}